\documentclass[proceedings]{JHEP3}
\usepackage{amsfonts}
\usepackage{amssymb}
\usepackage{amsfonts}
\usepackage{amsmath}
\usepackage{epsfig}

\newcommand{\be}{\begin{equation}}
\newcommand{\en}{\end{equation}}
\newcommand{\bea}{\begin{eqnarray}}
\newcommand{\ena}{\end{eqnarray}}
\newcommand{\beano}{\begin{eqnarray*}}
\newcommand{\enano}{\end{eqnarray*}}
\newcommand{\bee}{\begin{enumerate}}
\newcommand{\ene}{\end{enumerate}}

\conference{From Pseudo-bosons to Pseudo-Hermiticity}
\abstract{We consider the special type of pseudo-bosonic systems that can be mapped to standard bosons by means of generalized 
Bogoliubov transformation and demonstrate that a pseudo-Hermitian systems can be obtained from them by means of a second subsequent 
Bogoliubov transformation. We employ these operators in a simple model and study three different types of scenarios for the constraints
on the model parameters giving rise to a Hermitian system, a pseudo-Hermitian system in which the second the Bogoliubov transformations 
is equivalent to the associated Dyson map and one in which we obtain D-quasi bases.}

\title{From Pseudo-bosons to Pseudo-Hermiticity via multiple generalized
Bogoliubov transformations}
\author{Fabio Bagarello$^{\circ}$ and Andreas Fring$^\bullet$ \\
\noindent $^\circ$ Dipartimento di Energia, Ingegneria dell'Informazione e
Modelli Matematici, \\
$\,\,$ Facolt\`{a} di Ingegneria, Universit\`{a} di Palermo, I-90128
Palermo, Italy and \\
$\,\,$ INFN, Sezione di Torino, Italy\\
$^\bullet$ Department of Mathematics, City University London,\\
$\,\,$ Northampton Square, London EC1V 0HB, UK\\
$\,\,$ E-mail: fabio.bagarello@unipa.it, a.fring@city.ac.uk}

\input{tcilatex}
\begin{document}

\section{Introduction}

The basic principles for a consistent time-independent quantum mechanical
treatment of quasi/pseudo-Hermitian and/or $\mathcal{PT}$-symmetric
Hamiltonians are well understood theoretically and by now widely accepted,
see e.g. \cite{Bender:1998ke,Benderrev,Alirev,BFGJ} for an overview. In
addition, many experiments have been carried out to confirm the key findings
of this approach and to make further predictions, see e.g. \cite%
{Muss,MatMakris,Guo}.

A central element in such considerations is the construction of the
so-called Dyson map \cite{Dyson} that adjointly transforms a non-Hermitian
Hamiltonian to a Hermitian isospectral counterpart. Subsequently this map
can be used to manufacture a new metric operator for the physical Hilbert
space. This programme is conceptually straightforward, but it remains a
technical challenge even for simple examples \cite{Bender:2004sa,Mosta}.
Nonetheless, it has been carried out successfully for many concrete models 
\cite{sinha2002iso,fringfaria,ACIso,MGH,PEGAAF2}. Alternatively, one may
also attempt to transform a model's more basic constituents, such as Lie
algebraic \cite{PEGAAF2} or bosonic \cite{Swanson,PEGAAF2} building blocks
by different types of transformations. By viewing the adjoint map as a
generalized Bogoliubov transformation we pursue here a combined approach to
achieve this goal. In \cite{FabAnd} we found that when the constraint on the
target Hamiltonian to be Hermitian is relaxed, the generalized Bogoliubov
transformations still lead to systems with $\mathcal{D}$-pseudo-bosons, see
e.g. \cite{adjbook} for an overview, as their central constituents. We will
also consider such a scenario here by maintaining the structure of a
two-fold Bogoliubov transformations, where one of them is making up the
pseudo-bosons and the other is taken to be equivalent to an adjoint map. We
will also study the situation in which the constraint on the target objects
is relaxed.

Our manuscript is organized as follows: In section 2 we define our doubly
Bogoliubov transformed pseudo-bosons. In section 3 we study a Hamiltonian
built from the pseudo-bosonic number operator in various constraint
settings. We state our conclusions and an outlook in section 4.

\section{Adjointly transformed pseudo-bosons}

We consider here systems whose basic constituents are pseudo-bosonic
creation and destruction operators $c$ and $d$, respectively. These
operators satisfy the standard canonical commutation relations $[d,c]=%
\mathbb{I}$, but they are not mutually Hermitian, i.e. $d^{\dagger }\neq c$.
In general Hamiltonian systems comprised out of these operators will
therefore be non-Hermitian. Motivated by the success of
pseudo/quasi-Hermitian system we address here the question of whether and
how these operators can be mapped adjointly into a pair of almost mutually
Hermitian canonical operators, as such a map could be utilized to restore
the Hermiticity of the entire Hamiltonian system. Hence we seek to solve 
\begin{equation}
\eta \left( 
\begin{array}{l}
d \\ 
c%
\end{array}%
\right) \eta ^{-1}=\left( 
\begin{array}{l}
e \\ 
f%
\end{array}%
\right) ,\qquad \text{with }e^{\dagger }=\varkappa f,\varkappa \in \mathbb{R}%
,  \label{phpb}
\end{equation}%
for $\eta $. This general problem may be tackled in various generic manners
depending on the type of pseudo-bosons considered. Here we choose a specific
realization by taking the pseudo-bosons to be related to the standard
canonical creation and annihilation operators, $a$ and $a^{\dagger }$ with $%
[a,a^{\dagger }]=\mathbb{I}$, by means of a generalized Bogoliubov
transformation $T(\alpha ,\beta ,\gamma ,\delta )$, see \cite%
{Bogo,GenBog,Swanson,PEGAAF2,FabAnd},%
\begin{equation}
\left( 
\begin{array}{l}
d \\ 
c%
\end{array}%
\right) =T(\alpha ,\beta ,\gamma ,\delta )\left( 
\begin{array}{l}
a \\ 
a^{\dagger }%
\end{array}%
\right) .\quad  \label{pbosons}
\end{equation}%
Notice that other possibilities exist and see \cite{FabAnd} for a detailed
discussion on what the choice (\ref{pbosons}) entails. For the matrix $T$ we
assume the form%
\begin{equation}
T(\alpha ,\beta ,\gamma ,\delta )=\left( 
\begin{array}{cc}
\beta & -\delta \\ 
-\alpha & \gamma%
\end{array}%
\right) ,\qquad \det T(\alpha ,\beta ,\gamma ,\delta )=1,\qquad \alpha
,\beta ,\gamma ,\delta \in \mathbb{C},  \label{detT}
\end{equation}%
such that $d=\beta a-\delta a^{\dagger }$ and $c=-\alpha a+\gamma a^{\dagger
}$. Whereas $a$ and $a^{\dagger }$ are mutually Hermitian, $\left( a\right)
^{\dagger }=$ $a^{\dagger }$, this is obviously not the case for the
pseudo-bosons, unless $\beta =\gamma ^{\ast }$ and $\alpha =\delta ^{\ast }$.

Next we assume that the adjoint action on the standard bosons can also be
realized by a generalized Bogoliubov transformation%
\begin{equation}
\eta \left( 
\begin{array}{l}
a \\ 
a^{\dagger }%
\end{array}%
\right) \eta ^{-1}=T(\hat{\alpha},\hat{\beta},\hat{\gamma},\hat{\delta}%
)\left( 
\begin{array}{l}
a \\ 
a^{\dagger }%
\end{array}%
\right) \mathbf{.}  \label{B2}
\end{equation}%
This is indeed possible, taking for instance $\eta $ to be the positive
Hermitian operator%
\begin{equation}
\eta =\exp (\varepsilon a^{\dagger }a+\nu aa+\nu ^{\ast }a^{\dagger
}a^{\dagger }),\qquad \text{with }\varepsilon ^{2}\geq 4\left\vert \nu
\right\vert ^{2},
\end{equation}%
we find that the parameter $\hat{\alpha},\hat{\beta},\hat{\gamma},\hat{\delta%
}$ and $\varepsilon ,\nu $ are related as 
\begin{equation}
\hat{\alpha}=-2\frac{\nu }{\theta }\sinh \theta ,\quad \hat{\beta}=\cosh
\theta -\frac{\varepsilon }{\theta }\sinh \theta ,\quad \hat{\gamma}=\cosh
\theta +\frac{\varepsilon }{\theta }\sinh \theta ,\quad \hat{\delta}=2\frac{%
\nu ^{\ast }}{\theta }\sinh \theta ,  \label{sol}
\end{equation}%
where $\theta :=\sqrt{\varepsilon ^{2}-4\left\vert \nu \right\vert ^{2}}$.
The assumption $\det T(\hat{\alpha},\hat{\beta},\hat{\gamma},\hat{\delta})=1$
holds without any further constraint.

We may now solve (\ref{phpb}) by computing%
\begin{equation}
T(\alpha ,\beta ,\gamma ,\delta )\cdot T(\hat{\alpha},\hat{\beta},\hat{\gamma%
},\hat{\delta})\left( 
\begin{array}{l}
a \\ 
a^{\dagger }%
\end{array}%
\right) \mathbf{=}\eta T(\alpha ,\beta ,\gamma ,\delta )\left( 
\begin{array}{l}
a \\ 
a^{\dagger }%
\end{array}%
\right) \eta ^{-1}=\eta \left( 
\begin{array}{l}
d \\ 
c%
\end{array}%
\right) \eta ^{-1},  \label{TT}
\end{equation}%
where we used that evidently $[\eta ,T]=0$ and (\ref{pbosons}). From the
matrix multiplication on the left hand side and (\ref{phpb}) we obtain%
\begin{equation}
T(\tilde{\alpha},\tilde{\beta},\tilde{\gamma},\tilde{\delta})\left( 
\begin{array}{l}
a \\ 
a^{\dagger }%
\end{array}%
\right) =\left( 
\begin{array}{l}
e \\ 
f%
\end{array}%
\right) ,  \label{Ttil}
\end{equation}%
with 
\begin{equation}
\tilde{\alpha}=\alpha \hat{\beta}+\gamma \hat{\alpha},\quad \tilde{\beta}%
=\beta \hat{\beta}+\delta \hat{\alpha},\quad \tilde{\gamma}=\alpha \hat{%
\delta}+\gamma \hat{\gamma},\quad \tilde{\delta}=\beta \hat{\delta}+\delta 
\hat{\gamma}.
\end{equation}%
Since the determinants of $T(\alpha ,\beta ,\gamma ,\delta )$ and $T(\hat{%
\alpha},\hat{\beta},\hat{\gamma},\hat{\delta})$ are $1$, we also have $\det
T(\tilde{\alpha},\tilde{\beta},\tilde{\gamma},\tilde{\delta})=1$. Depending
now on the constraints imposed on the Bogoliubov transformation parameters $%
\alpha ,\beta ,\gamma ,\delta $ and those entering from the adjoint action $%
\varepsilon ,\nu $ we obtain different types of scenarios, which we now
investigate for a concrete model.

\section{Pseudo-bosonic Hamiltonians}

We consider here a system described by a Hamiltonian consisting of the
pseudo-bosonic number operator $N=cd$ of the type studied in \cite{FabAnd} 
\begin{equation}
\mathcal{H}(d,c)=\hbar \omega \left( cd+\frac{1}{2}\right) ,  \label{1}
\end{equation}%
where in comparison we re-introduced the standard angular frequency $\omega
\in \mathbb{R}$ and the reduced Planck constant $\hbar $. Assuming here that
the pseudo-bosons are generated by a generalized Bogoliubov transformation
as specified in (\ref{pbosons}) the Hamiltonian in (\ref{1}) acquires the
form of a Swanson Hamiltonian \cite{Swanson} 
\begin{equation}
\mathcal{H}(a,a^{\dagger })=\hbar \omega \left( \alpha \delta aa^{\dagger
}+\beta \gamma a^{\dagger }a-\alpha \beta aa-\gamma \delta a^{\dagger
}a^{\dagger }+\frac{1}{2}\right) .  \label{HH}
\end{equation}%
Evidently the Hamiltonian $\mathcal{H}$ is non-Hermitian whenever $\alpha
\delta \neq \alpha ^{\ast }\delta ^{\ast }$ or $\beta \gamma \neq \beta
^{\ast }\gamma ^{\ast }$ or $\alpha \beta \neq \gamma ^{\ast }\delta ^{\ast
} $. We note that the transformation $\alpha \leftrightarrow \delta ^{\ast }$%
, $\beta \leftrightarrow \gamma ^{\ast }$ maps $\mathcal{H\rightarrow H}%
^{\dagger }$, which implies that our Hamiltonian becomes Hermitian when this
transformation becomes a symmetry. Let us now consider various cases for
possible constraints on the parameters involved.

\subsection{Hermitian constraint}

It is instructive to consider at first the simplest special scenario with $%
\alpha ,\beta ,\gamma ,\delta \in \mathbb{R}$ and the additional constraint $%
\alpha \beta =\gamma \delta $. In this case the Hamiltonian $\mathcal{H}$ in
(\ref{HH}) evidently becomes Hermitian, so that we do not require the
similarity transformation (\ref{phpb}) to achieve this, such that $\eta =%
\mathbb{I}$. We may solve the constraint together with the restrictions on
the determinant of the Bogoliubov transformation in (\ref{detT}) for two of
the constants, e.g.%
\begin{equation}
\beta =\frac{\gamma }{\gamma ^{2}-\alpha ^{2}},\qquad \text{and\qquad }%
\delta =\frac{\alpha }{\gamma ^{2}-\alpha ^{2}}.  \label{bd}
\end{equation}%
Notice that although $\mathcal{H}=\mathcal{H}^{\dagger }$, we still maintain
the pseudo-bosonic property $d^{\dagger }\neq c$ with%
\begin{equation}
d^{\dagger }=\frac{1}{\gamma ^{2}-\alpha ^{2}}c.
\end{equation}%
Using now the common representation for the canonical creation and
annihilation operators%
\begin{equation}
a(m,\omega ):=\frac{1}{\sqrt{2\hbar }}\left( \sqrt{m\omega }x+\frac{i}{\sqrt{%
m\omega }}p\right) ,\qquad a^{\dagger }(m,\omega ):=\frac{1}{\sqrt{2\hbar }}%
\left( \sqrt{m\omega }x-\frac{i}{\sqrt{m\omega }}p\right) ,  \label{cc}
\end{equation}%
in terms of the canonical coordinate and momentum operator $x$ and $p$,
obeying $[x,p]=i\hbar $, the Hamiltonian (\ref{1}) acquires the form of the
standard harmonic oscillator Hamiltonian%
\begin{equation}
\mathcal{H}=h=\hbar \omega \left[ a^{\dagger }(M,\omega )a(M,\omega )+\frac{1%
}{2}\right] =\frac{1}{2M}p^{2}+\frac{M\omega ^{2}}{2}x^{2},  \label{Harm}
\end{equation}%
albeit with a modified mass%
\begin{equation}
M=m\frac{\gamma -\alpha }{\gamma +\alpha }=m\frac{\beta -\delta }{\beta
+\delta }.
\end{equation}%
In order to ensure the mass to be physical, that is positive, we require $%
\gamma >\alpha >0$ and $\beta >\delta >0$, which together with $\alpha \beta
=\gamma \delta $ are precisely the constraints encountered in \cite{FabAnd}
for this situation as a requirement for the eigenfunctions of $\mathcal{H}$
to be square-integrable. Given the well-known form for the normalized
eigenfunctions for the Hamiltonian (\ref{Harm}) in terms of Hermite
polynomials $H_{n}$ 
\begin{equation}
\phi _{n}(x)=\frac{1}{\sqrt{n!2^{n}}}\left( \frac{M\omega }{\pi }\right)
^{1/4}e^{-\omega Mx^{2}/2}H_{n}(M\omega x),
\end{equation}%
these two requirements are therefore the same. In other words, square
integrability and positivity of the mass become synonymous, depending both
explicitly on the model parameters.

\subsection{Pseudo-Hermitian constraint}

Let us now relax the constraint $\alpha \beta =\gamma \delta $ and only
assume $\alpha ,\beta ,\gamma ,\delta \in \mathbb{R}$. We carry out an
analysis using standard techniques developed in the context of
pseudo/quasi-Hermitian and/or $\mathcal{PT}$-symmetric quantum mechanics as
outlined in \cite{Urubu,Bender:1998ke,Benderrev,Alirev}. To establish our
notation we briefly recall the key formulae.

Given two isospectral Hamiltonians one of which is Hermitian $h=h^{\dagger }$
and the other is not $\mathcal{H}\neq \mathcal{H}^{\dagger }$, related by a
similarity transformation%
\begin{equation}
h=\eta \mathcal{H}\eta ^{-1},  \label{hH}
\end{equation}%
the corresponding eigenstates for the eigenvalue equations 
\begin{equation}
\mathcal{H}\left\vert \psi \right\rangle =E\left\vert \psi \right\rangle
\qquad \text{and\qquad }h\left\vert \phi \right\rangle =E\left\vert \phi
\right\rangle ,
\end{equation}%
are related as%
\begin{equation}
\left\vert \psi \right\rangle =\eta ^{-1}\left\vert \phi \right\rangle .
\end{equation}%
The expectation values for any observable $\mathcal{O}$ and $\mathit{o}$, in
a non-Hermitian and corresponding Hermitian counterpart, respectively, are
related as $\left\langle \psi \right\vert \mathcal{O}\left\vert \psi
\right\rangle _{\eta }=$ $\left\langle \phi \right\vert \mathit{o}\left\vert
\phi \right\rangle $. Here the inner product is defined as $\left\langle
\psi \right\vert \left. \psi \right\rangle _{\eta }:=\left\langle \psi
\right\vert \eta ^{\dagger }\eta \left\vert \psi \right\rangle $ where the
positive operator $\eta ^{\dagger }\eta $ plays the role of the metric.

Let us now use the above formulae to carry out the similarity transformation
for the non-Hermitian Hamiltonian (\ref{1}) to a Hermitian isospectral
counterpart 
\begin{equation}
h=\eta \mathcal{H}(d,c)\eta ^{-1}=\hbar \omega \left( \eta c\eta ^{-1}\eta
d\eta ^{-1}+\frac{1}{2}\right) =\hbar \omega \left( fe+\frac{1}{2}\right) .
\label{h}
\end{equation}%
The transformed Hamiltonian is Hermitian when $e^{\dagger }=\varkappa f$
with $\varkappa \in \mathbb{R}$, which according to (\ref{Ttil}) can be
achieved by imposing the two additional constraints%
\begin{equation}
\tilde{\delta}=\varkappa \tilde{\alpha},\quad \text{and\quad }\tilde{\beta}%
=\varkappa \tilde{\gamma},  \label{xi}
\end{equation}%
with $\nu \in \mathbb{R}$. Eliminating $\varkappa $ by using the explicit
expressions in (\ref{sol}), these two equations reduce to 
\begin{equation}
\frac{\tanh 2\theta }{\theta }=\frac{\alpha \beta -\gamma \delta }{%
\varepsilon (\alpha \beta +\gamma \delta )+2\nu (\beta \gamma +\alpha \delta
)}.  \label{const}
\end{equation}%
Assuming the parameters $\alpha ,\beta ,\gamma ,\delta $ to be model
specific, and therefore fixed, the two additional parameters $\varepsilon $
and $\nu $ that entered through the second generalized Bogoliubov
transformation are constrained by (\ref{const}). Thus there is only one free
parameter left, which reflects the typical ambiguity present in
pseudo-Hermitian systems, see \cite{Urubu} for the Swanson model at hand.
Parameterizing one of the free parameters in terms of the other as $\nu
=\lambda \varepsilon /2$, we obtain $\varepsilon $ as a function of the new
variable $\lambda $ 
\begin{equation}
\varepsilon (\lambda )=\frac{1}{2\sqrt{1-\lambda ^{2}}}\func{arctanh}\left[ 
\frac{(\alpha \beta -\gamma \delta )\sqrt{1-\lambda ^{2}}}{(\alpha \beta
+\gamma \delta )+\lambda (\beta \gamma +\alpha \delta )}\right] ,\qquad
\lambda :=\frac{2\nu }{\varepsilon }.  \label{eps}
\end{equation}%
The restrictions of the interval in which $\lambda $ is taken results from
demanding $\theta \in \mathbb{R}$. Recalling that $\varepsilon $ is real, we
restrict the argument of the $\func{arctanh}$ to be bounded by $\pm 1$.
Choosing for definiteness as specific ordering $0<\alpha /\gamma <\delta
/\beta <1$, we need to restrict $\lambda $ further to be in the disconnected
intervals%
\begin{equation}
-\infty <\lambda <-\frac{2\beta \delta }{\beta ^{2}+\delta ^{2}},\qquad 
\text{or\qquad }-\frac{2\alpha \gamma }{\alpha ^{2}+\gamma ^{2}}<\lambda
<\infty ,  \label{lint}
\end{equation}%
for $\varepsilon $ to remain real. Notice that the restrictions on the model
parameters are just needed to ensure that the constraints (\ref{lint})
become unique. The intervals are only connected for $\alpha \beta =\gamma
\delta $, which corresponds to the Hermitian case discussed in the previous
subsection.

Given the above transformations we can of course express our isospectral
Hamiltonian (\ref{h}) also in terms of the standard bosonic operators%
\begin{equation}
h(m,\omega )=\hbar \omega \left[ \mu _{a^{\dagger }a}a^{\dagger }(m,\omega
)a(m,\omega )-\mu _{aa}a(m,\omega )a(m,\omega )-\mu _{a^{\dagger }a^{\dagger
}}a^{\dagger }(m,\omega )a^{\dagger }(m,\omega )+\mu _{0}\right] ,
\end{equation}%
with coefficients%
\begin{equation}
\mu _{a^{\dagger }a}=\tilde{\alpha}\tilde{\delta}+\tilde{\beta}\tilde{\gamma}%
,\quad \mu _{aa}=\tilde{\alpha}\tilde{\beta},\quad \mu _{a^{\dagger
}a^{\dagger }}=\tilde{\gamma}\tilde{\delta},\quad \text{and}\quad \mu _{0}=%
\frac{1}{2}+\tilde{\alpha}\tilde{\delta}.
\end{equation}%
The constraint (\ref{xi}) guarantees that $\mu _{aa}=$\ $\mu _{a^{\dagger
}a^{\dagger }}$, such that together with $\mu _{0},\mu _{a^{\dagger }a}\in 
\mathbb{R}$ the Hamiltonian $h$ becomes Hermitian. In fact these constraints
are familiar from a general treatment of the Swanson model \cite{MGH}, which
is a special case of the more general model studied in \cite{PEGAAF2} and
precisely agrees when matching the constants appropriately. However, whereas
in \cite{MGH,PEGAAF2} the constraints resulted from an analysis of the
Hamiltonian, they emerge here as the combination of two constraints on their
more basic pseudo-bosonic constituents.

Just as the Hermitian case in the previous subsection, when implementing the
constraints the Hamiltonian $h$ can be brought into the form of a harmonic
oscillator 
\begin{equation}
h(\hat{M},\omega )=\hbar \omega \left[ a^{\dagger }(\hat{M},\omega )a(\hat{M}%
,\omega )+\frac{1}{2}\right] =\frac{1}{2\hat{M}}p^{2}+\frac{\hat{M}\omega
^{2}}{2}x^{2},  \label{HHH}
\end{equation}%
again with modified mass 
\begin{equation}
\hat{M}=\frac{m}{(\tilde{\alpha}+\tilde{\gamma})(\tilde{\beta}+\tilde{\delta}%
)}.
\end{equation}%
Notice that using the constraint (\ref{xi}) we only need to take $m\varkappa
>0$ in order to ensure that $\hat{M}$ is positive, when $\tilde{\alpha},%
\tilde{\beta},\tilde{\gamma},\tilde{\delta}\in \mathbb{R}.$ Viewing $\alpha
,\beta ,\gamma ,\delta $ as model defining parameters, it follows from (\ref%
{sol}), (\ref{xi}) and (\ref{eps}), that for a specific model we may regard $%
\hat{M}$ and $\omega $ as functions of the single parameter $\lambda $. For
some specific choices we show in figure 1 the modified mass as a function of 
$\lambda $. We observe that it is positive and according to (\ref{lint}) not
defined in the specified interval for $\lambda $.

We recall that $\lambda $ is not a model dependent parameter as it simply
entered through the adjoint map labeling infinitely many pseudo-Hermitian
counterparts to $\mathcal{H}(d,c)$. Any of the theories respecting the
constraint (\ref{lint}) is well defined. As noted in \cite{MGH}, the
theories for $\lambda =0,\pm 1$ are somewhat special as then some of the
auxiliary variables can be interpreted directly as the number operator, the
coordinate or the momentum. In these cases we find%
\begin{equation}
\hat{M}(\lambda =0)=\frac{m}{\left( \sqrt{\alpha \delta }+\sqrt{\beta \gamma 
}\right) ^{2}},\quad \text{and\quad }\hat{M}(\lambda =\pm 1)=\frac{m}{\left[
\left( \gamma \pm \alpha \right) \left( \beta \pm \delta \right) \right]
^{\pm 1}},  \label{MM}
\end{equation}%
which are all well-defined for the ordering considered here. We may confirm
these expression for the values used in our examples in figure 1. For
instance, for the choice of parameters corresponding to the solid black line
we compute\ $\hat{M}(\lambda =0)=0.672$, $\hat{M}(\lambda =1)=0.518$ and $%
\hat{M}(\lambda =-1)=0.48$ numerically and also obtain the same values from
the explicit analytical expression (\ref{MM}).

\FIGURE{\epsfig{file=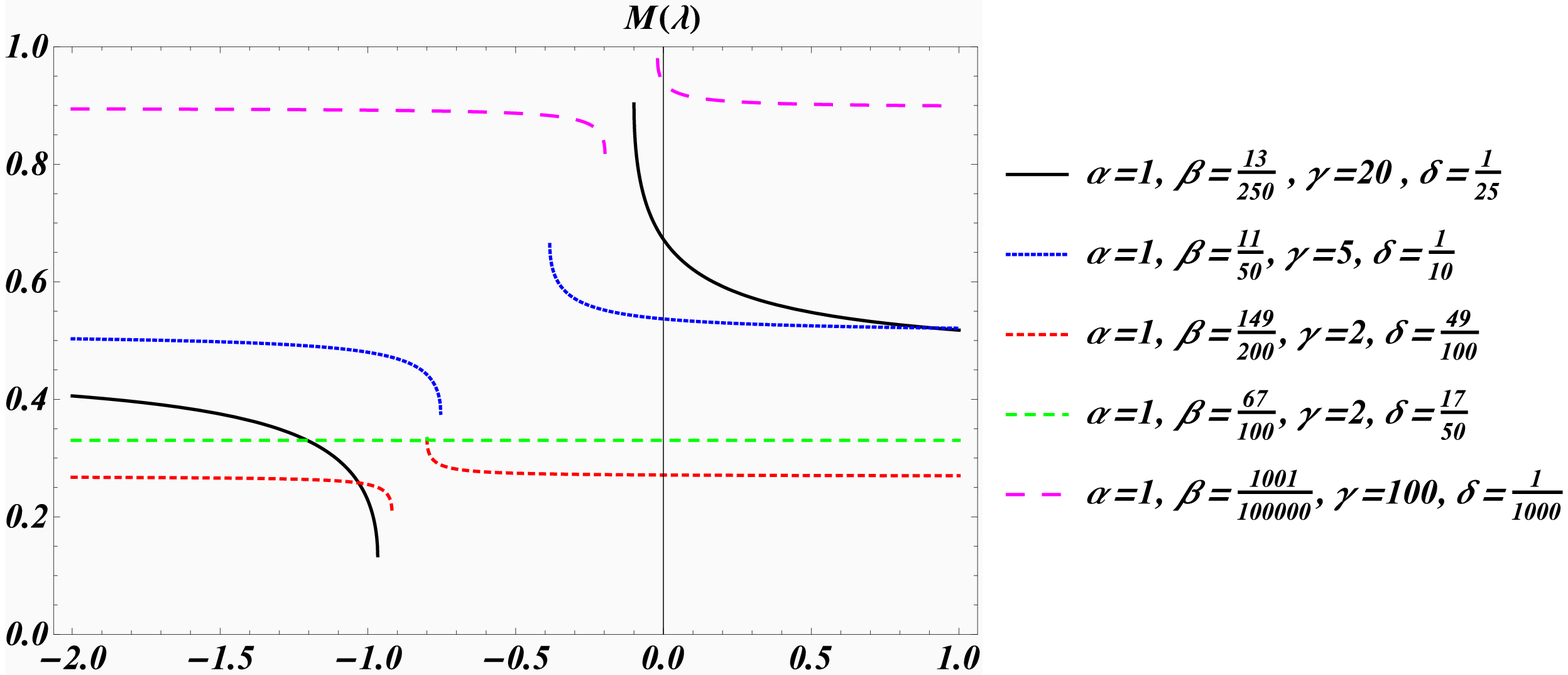,height=6.5cm}[h]  
        \caption{Modified mass $M$ as a function of $\lambda$ for $m=1$ with $\lambda \in \mathbb{R}\backslash (-0.9665,-0.0998)$, $\lambda \in \mathbb{R}\backslash
(-0.7534,-0.3846)$, $\lambda \in \mathbb{R}\backslash (-0.9182,-0.8)$, $\lambda \in \mathbb{R}\backslash (-0.80709,-0.8)$ and $\lambda \in \mathbb{R}\backslash (-0.1978,-0.0200)$.}
        \label{MassL}}

We expect to recover the Hermitian case by demanding either the position and
the momentum, the position and the number operator or the momentum and the
number operator to be observables. Indeed by equating any two of these
masses, i.e. $\hat{M}(0)=\hat{M}(\pm 1)$ or $\hat{M}(1)=\hat{M}(-1)$,
together with the constraint on the determinant of $T$ leads to the values
in (\ref{bd}) for two of the parameters.

\subsection{Adjoint constraint}

Let us now also relax the constraint (\ref{xi}), such that $e^{\dagger }\neq
\varkappa f$ and in addition allow $\alpha ,\beta ,\gamma ,\delta \in 
\mathbb{C}$. Whereas in the previous subsections the construction of the
eigenfunctions follows trivially from the harmonic oscillator realization in
terms of modified masses and frequencies, this is less obvious for this
setting. We therefore present the construction commencing by expressing the
pseudo-bosons $e$ and $f$ in position space. From (\ref{Ttil}) and (\ref{cc}%
) simply follows 
\begin{equation}
e=\frac{1}{\sqrt{2}}\left[ \left( \tilde{\beta}-\tilde{\delta}\right)
x+\left( \tilde{\beta}+\tilde{\delta}\right) \frac{d}{dx}\right] ,\quad f=%
\frac{1}{\sqrt{2}}\left[ \left( \tilde{\gamma}-\tilde{\alpha}\right)
x-\left( \tilde{\alpha}+\tilde{\gamma}\right) \frac{d}{dx}\right] ,
\end{equation}%
As we no longer need modified values for the mass and frequency we have set
here $m=\omega =\hbar =1$ for simplicity. Hence, following \cite{FabAnd},
the vacua of $e$ and $f^{\dagger }$ are 
\begin{equation}
\varphi _{0}(x)=N_{\varphi }e^{-\frac{1}{2}\,x^{2}\,\frac{\tilde{\beta}-%
\tilde{\delta}}{\tilde{\beta}+\tilde{\delta}}},\quad \quad \text{and\quad
\quad }\Psi _{0}(x)=N_{\Psi }e^{-\frac{1}{2}\,x^{2}\,\frac{\tilde{\gamma}%
^{\ast }-\tilde{\alpha}^{\ast }}{\tilde{\gamma}^{\ast }+\tilde{\alpha}^{\ast
}}},  \label{f03}
\end{equation}%
respectively, where $N_{\varphi }$ and $N_{\Psi }$ are suitable
normalization constants to be specified further below. Naturally we require 
\begin{equation}
\func{Re}\left( \frac{\tilde{\beta}-\tilde{\delta}}{\tilde{\beta}+\tilde{%
\delta}}\right) >0,\quad \quad \text{and\quad \quad }\func{Re}\left( \frac{%
\tilde{\gamma}-\tilde{\alpha}}{\tilde{\gamma}+\tilde{\alpha}}\right) >0,
\label{f04}
\end{equation}%
to ensure the square-integrability for both of these functions. In complete
analogy to \cite{FabAnd} we further construct the functions 
\begin{eqnarray}
\varphi _{n}(x) &=&\frac{f^{n}}{\sqrt{n!}}\varphi _{0}(x)=\frac{N_{\varphi }%
}{\sqrt{n!2^{n}}}\left( \frac{\tilde{\alpha}+\tilde{\gamma}}{\tilde{\beta}+%
\tilde{\delta}}\right) ^{n/2}H_{n}\left[ \frac{x}{\sqrt{(\tilde{\alpha}+%
\tilde{\gamma})(\tilde{\beta}+\tilde{\delta})}}\right] \,e^{-\frac{1}{2}%
\,x^{2}\,\frac{\tilde{\beta}-\tilde{\delta}}{\tilde{\beta}+\tilde{\delta}}},
\\
\Psi _{n}(x) &=&\frac{{e^{\dagger }}^{n}}{\sqrt{n!}}\Psi _{0}(x)=\frac{%
N_{\Psi }}{\sqrt{n!2^{n}}}\left( \frac{\tilde{\beta}^{\ast }+\tilde{\delta}%
^{\ast }}{\tilde{\alpha}^{\ast }+\tilde{\gamma}^{\ast }}\right) ^{n/2}H_{n}%
\left[ \frac{x}{\sqrt{(\tilde{\alpha}^{\ast }+\tilde{\gamma}^{\ast })(\tilde{%
\beta}^{\ast }+\tilde{\delta}^{\ast })}}\right] \,e^{-\frac{1}{2}\,x^{2}\,%
\frac{\tilde{\gamma}^{\ast }-\tilde{\alpha}^{\ast }}{\tilde{\gamma}^{\ast }+%
\tilde{\alpha}^{\ast }}}.  \notag
\end{eqnarray}%
from a repeated action of $f$ and ${e^{\dagger }}$ on the corresponding
ground states in (\ref{f03}) for $n\geq 0$. When the constraint (\ref{f04})
holds these functions are square-integrable and can be used to define the
sets $\mathcal{F}_{\varphi }:=\{\varphi _{n},n\geq 0\}$ and $\mathcal{F}%
_{\Psi }:=\{\Psi _{n},n\geq 0\}$. Applying here what was proven in \cite%
{FabAnd}, we deduce that $\mathcal{F}_{\varphi }$ and $\mathcal{F}_{\Psi }$
form biorthogonal bases for $\mathcal{H}$ when $\tilde{\alpha}\,\tilde{\beta}%
=\tilde{\gamma}\,\tilde{\delta}$, i.e. we have 
\begin{equation}
\chi (x)=\sum_{n=0}^{\infty }\left\langle \Psi _{n},\chi \right\rangle
\varphi _{n}(x)=\sum_{n=0}^{\infty }\left\langle \varphi _{n},\chi
\right\rangle \Psi _{n}(x),
\end{equation}%
for all $\chi (x)\in \mathcal{L}^{2}(\mathbb{R)}$. Notice that $\tilde{\alpha%
}\,\tilde{\beta}=\tilde{\gamma}\,\tilde{\delta}$ is simply the constraint (%
\ref{xi}) when eliminating $\varkappa $, such that $h$ becomes self-adjoint.
When this constraint is relaxed, i.e. $\tilde{\alpha}\,\tilde{\beta}\neq 
\tilde{\gamma}\,\tilde{\delta}$, the two sets still form ${\mathcal{D}}$%
-quasi bases, i.e. we have 
\begin{equation}
\left\langle \chi ,\xi \right\rangle =\sum_{n=0}^{\infty }\left\langle \chi
,\Psi _{n}\right\rangle \left\langle \varphi _{n},\xi \right\rangle
=\sum_{n=0}^{\infty }\left\langle \chi ,\varphi _{n}\right\rangle
\left\langle \Psi _{n},\xi \right\rangle ,
\end{equation}%
for all $\chi ,\xi \in {\mathcal{D}}$, a dense subset of $\mathcal{L}^{2}(%
\mathbb{R)}$ defined as follows: 
\begin{equation}
{\mathcal{D}}=\left\{ \chi (x)\in \mathcal{L}^{2}(\mathbb{R)}:\,e^{\frac{1}{2%
}x^{2}|\tilde{\alpha}\tilde{\beta}-\tilde{\gamma}\tilde{\delta}|}\chi (x)%
\mathbb{\in }\mathcal{L}^{2}(\mathbb{R)}\right\} .
\end{equation}

It is possible to verify that each $\Psi _{n}(x)$ is in the domain of $\eta $
while each $\varphi _{n}(x)$ is in the domain of $\eta ^{-1}$, that is $\Psi
_{n}(x)\in D(\eta )$ and $\varphi _{n}(x)\in D(\eta ^{-1})$. We prove this
as follows: By the definition of $\varphi _{n}$ we know that $\varphi _{n+1}=%
\frac{1}{\sqrt{n+1}}\,f\varphi _{n}$, for all $n\geq 0$. Since $\varphi
_{k}\in \mathcal{L}^{2}(\mathbb{R)}$ for all $k$, we conclude that $\varphi
_{n}\in D(f)$. Recalling now that $f=\eta c\eta ^{-1}$, it follows that $%
D(f)\subseteq D(\eta ^{-1})$. Hence, since $\varphi _{n}\in D(f)$ this means 
$\varphi _{n}\in D(\eta ^{-1})$. Similarly we can check that each $\Psi
_{n}\in D(\eta )$. These facts are important, since they imply that both $%
\eta $ and $\eta ^{-1}$ are densely defined. In fact, $\eta ^{-1}$ is
defined on $\mathcal{L}_{\varphi }$, the linear span of the $\varphi _{n}$%
's, which is dense in $\mathcal{H}$ since $\mathcal{F}_{\varphi }$ is
complete (or, if $\tilde{\alpha}\,\tilde{\beta}=\tilde{\gamma}\,\tilde{\delta%
}$, is even a basis). Similarly, $\eta $ is defined on $\mathcal{L}_{\Psi }$%
, the linear span of the $\Psi _{n}$'s, which is dense in $\mathcal{H}$
since $\mathcal{F}_{\Psi }$ is complete (or, again, if $\tilde{\alpha}\,%
\tilde{\beta}=\tilde{\gamma}\,\tilde{\delta}$, is a basis).

More detail on this case may be found in \cite{FabAnd}.

\section{Conclusions}

We further investigated a particular type of pseudo-bosons that are
obtainable from generalized Bogoliubov transformations \cite{FabAnd}. We
apply on them a second generalized Bogoliubov transformation satisfying
certain properties and demand that it equals a particular adjoint map acting
on these operators. We employ these doubly transformed operators to build a
simple Hamiltonian consisting of these special pseudo-bosonic number
operators.

We impose constraints on the model parameters, which we gradually relax.
Choosing at first the model parameters in such a way that the Hamiltonian
becomes Hermitian we found that one requires further simple constraints on
the ordering of the model parameters in order to obtain a positive mass.
This requirement turned out to be the same as demanding square integrability
of the wave functions. As the next case we demand the adjoint map to be
equivalent to the Dyson map that achieves pseudo-Hermiticity. In this
setting, we obtain the typical scenario in pseudo-Hermitian systems namely a
whole ray of equivalent Hermitian Hamiltonians (\ref{HHH}) parametrized by a
non-model dependent quantity, $\lambda $ in our case, entering through the
similarity transformation. We found that $\lambda $ is always defined on two
disjoint intervals on the real line. In the excluded parameter regime the
mass becomes complex as a consequence of $\varepsilon (\lambda )$ in (\ref%
{eps}) becoming complex. Finally when relaxing all constraints we loose the
proper that $\mathcal{F}_{\varphi }$ and $\mathcal{F}_{\Psi }$ form
biorthogonal bases as in the Hermitian scenario, but we still obtain ${%
\mathcal{D}}$-quasi bases.

The main virtue of our construction lies in the reduction of the relevant
transformations to the more basic bosonic ingredients. Even though the
doubly Bogoliubov transformed objects are more restrictive when compared to
the most general treatment, they always select out a set of feasible models.
Naturally, they may be employed in other more complicated models, involving
for instance cubic \cite{PEGAAF} or higher order terms in its defining
Hamiltonian.\medskip

\noindent \textbf{Acknowledgments:} FB gratefully acknowledges financial
support from City University London, from the Universit\`{a} di Palermo, via
CORI 2014, Action D and from G.N.F.M.


\begin{thebibliography}{99}
\bibitem{Bender:1998ke} C.~M. Bender and S.~Boettcher, \newblock Real
Spectra in Non-Hermitian Hamiltonians Having PT Symmetry, \newblock Phys.
Rev. Lett. \textbf{80}, 5243--5246 (1998).

\bibitem{Benderrev} C.~M. Bender, \newblock Making sense of non-Hermitian
Hamiltonians, \newblock Rept. Prog. Phys. \textbf{70}, 947--1018 (2007).

\bibitem{Alirev} A.~Mostafazadeh, \newblock Pseudo-Hermitian Representation
of Quantum Mechanics, \newblock Int. J. Geom. Meth. Mod. Phys. \textbf{7},
1191--1306 (2010).

\bibitem{BFGJ} C.~M. Bender, A.~Fring, U.~Guenther, and H.~F. Jones, %
\newblock Special issue on quantum physics with non-Hermitian operators, %
\newblock Journal of Physics A: Mathematical and Theoretical \textbf{45}(1),
010201 (2012).

\bibitem{Muss} Z.~H. Musslimani, K.~G. Makris, R.~El-Ganainy, and D.~N.
Christodoulides, \newblock Optical Solitons in PT Periodic Potentials, %
\newblock Phys. Rev. Lett. \textbf{100}, 030402 (2008).

\bibitem{MatMakris} K.~G. Makris, R.~El-Ganainy, D.~N. Christodoulides, and
Z.~H. Musslimani, \newblock PT-symmetric optical lattices, \newblock Phys.
Rev. \textbf{A81}, 063807(10) (2010).

\bibitem{Guo} A.~Guo, G.~J. Salamo, D.~Duchesne, R.~Morandotti,
M.~Volatier-Ravat, V.~Aimez,  G.~A. Siviloglou, and D.~Christodoulides, %
\newblock Observation of PT-Symmetry Breaking in Complex Optical Potentials, %
\newblock Phys. Rev. Lett. \textbf{103}, 093902(4) (2009).

\bibitem{Dyson} F.~J. Dyson, \newblock Thermodynamic Behavior of an Ideal
Ferromagnet, \newblock Phys. Rev. \textbf{102}, 1230--1244 (1956).

\bibitem{Bender:2004sa} C.~M. Bender, D.~C. Brody, and H.~F. Jones, %
\newblock Extension of PT-symmetric quantum mechanics to quantum field
theory  with cubic interaction, \newblock Phys. Rev. \textbf{D70},
025001(19) (2004).

\bibitem{Mosta} A.~Mostafazadeh, \newblock {PT-symmetric cubic anharmonic
oscilator as a physical model}, \newblock J. Phys. \textbf{A38}, 6557--6570
(2005).

\bibitem{sinha2002iso} A.~Sinha and R.~Roychoudhury, \newblock Isospectral
partners of a complex PT-invariant potential, \newblock Physics Letters A 
\textbf{301}(3), 163--172 (2002).

\bibitem{fringfaria} C.~Figueira~de Morisson~Faria and A.~Fring, \newblock
{Time evolution of non-Hermitian Hamiltonian systems}, \newblock J. Phys. 
\textbf{A39}, 9269--9289 (2006).

\bibitem{ACIso} C.~Figueira~de Morisson~Faria and A.~Fring, \newblock %
Isospectral Hamiltonians from Moyal products, \newblock Czech. J. Phys. 
\textbf{56}, 899--908 (2006).

\bibitem{MGH} D.~P. Musumbu, H.~B. Geyer, and W.~D. Heiss, \newblock Choice
of a metric for the non-Hermitian oscillator, \newblock J. Phys. \textbf{A40}%
, F75--F80 (2007).

\bibitem{PEGAAF2} P.~E.~G. Assis and A.~Fring, \newblock Non-Hermitian
Hamiltonians of Lie algebraic type, \newblock J. Phys. \textbf{A42}, 015203
(23p) (2009).

\bibitem{Swanson} M.~S. Swanson, \newblock Transition elements for a
non-Hermitian quadratic Hamiltonian, \newblock J. Math. Phys. \textbf{45},
585--601 (2004).

\bibitem{FabAnd} F.~Bagarello and A.~Fring, \newblock Generalized Bogoliubov
transformations versus  $\mathcal{D}$-pseudo-bosons, \newblock Journal of
Mathematical Physics \textbf{56}(10), 103508 (2015).

\bibitem{adjbook} F.~Bagarello, \newblock Deformed canonical
(anti-)commutation relations and non-hermitian  hamiltonians, \newblock in 
\emph{Non-Hermitian operators in quantum physics: Mathematical  aspects}, F.
Bagarello, J. P. Gazeau, F. H. Szafraniec and M. Znojil, Eds.,  (John Wiley
and Sons, New Jersey) (2015).

\bibitem{Bogo} N.~N. Bogolyubov, \newblock {On a new method in the theory of
superconductivity}, \newblock Nuovo Cim. \textbf{7}, 794--805 (1958).

\bibitem{GenBog} F.~Hong-Yi and J.~Vanderlinde, \newblock {Generalized
Bogolyubov transformation}, \newblock J. Phys. \textbf{A23}, L1113--L1117
(1990).

\bibitem{Urubu} F.~G. Scholtz, H.~B. Geyer, and F.~Hahne, \newblock %
Quasi-Hermitian Operators in Quantum Mechanics and the Variational 
Principle, \newblock Ann. Phys. \textbf{213}, 74--101 (1992).

\bibitem{PEGAAF} P.~E.~G. Assis and A.~Fring, \newblock Metrics and
isospectral partners for the most generic cubic  $\mathcal{PT}$-symmetric
non-Hermitian Hamiltonian, \newblock J. Phys. \textbf{A41}, 244001 (2008).
\end{thebibliography}
\end{document}